\documentclass[12pt,a4wide]{article}
\usepackage{amsmath,amssymb,amsfonts}

\newcommand{\eqa}{\begin{eqnarray}}
\newcommand{\ena}{\end{eqnarray}}
\newcommand{\topstar}[1]{\setlength{\unitlength}{1mm}
\begin{picture}(2,0)(-1,-1.4)
   \put(0,0){\makebox(0,0){$#1$}}
   \put(0,2.4){\makebox(0,0){\mbox{\tiny$\star$}}}
\end{picture}}
\begin{document}
\begin{center}
{\large {\bf Metrically Stationary, Axially Symmetric, Isolated Systems in 
Quasi-Metric Gravity}}
\end{center}
\begin{center}
Dag {\O}stvang \\
{\em Department of Physics, Norwegian University of Science and Technology
(NTNU) \\
N-7491 Trondheim, Norway}
\end{center}
\begin{abstract}
The gravitational field exterior respectively interior to an axially 
symmetric, metrically stationary, isolated spinning source made of perfect 
fluid is examined within the quasi-metric framework. (A metrically stationary 
system is defined as a system which is stationary except for the direct effects
of the global cosmic expansion on the space-time geometry.) Field equations 
are set up and an attempt is made to find an approximate series solution for 
the exterior part. However, the result is that no stationary solution 
corresponding to a spinning source can exist when considering terms beyond a 
certain order in small quantities. That is, except for metrically static 
systems, axially symmetric systems must necessarily be non-stationary in 
quasi-metric gravity. However, sufficiently weak, axially symmetric 
gravitational fields associated with slowly rotating sources, may still be 
considered as stationary to an excellent approximation.

Thus a truncated, approximately stationary solution is found for the exterior 
field. To lowest order in small quantities, the gravitomagnetic part of the 
found metric family corresponds with the Kerr metric in the metric 
approximation. On the other hand, the gravitoelectric part of the found metric 
family also includes a tidal term characterized by the free quadrupole-moment 
parameter $J_2$ describing the effect of source deformation due to the 
rotation. This term has no counterpart in the Kerr metric. Finally, the 
geodetic effect for a gyroscope in orbit is calculated. There is a correction 
term, unfortunately barely too small to be detectable by Gravity Probe B, to 
the standard expression.
\end{abstract}
\topmargin 0pt
\oddsidemargin 5mm
\section{Introduction} 
Sources of gravitation do as a rule rotate. The rotation should in itself
gravitate and thus affect the associated gravitational fields. In General 
Relativity (GR) this is well illustrated by the Kerr metric describing the 
gravitational field outside a spinning black hole. More generally, for cases
where no exact solutions exist, it is possible to find numerical solutions of 
the full Einstein equations, both interior and exterior to a stationary 
spinning source made of perfect fluid with a prescribed equation of state. 
Besides, on a more analytical level, weak field and slow angular velocity 
approximations are useful to show the dominant effects of rotation on 
gravitational fields. Such approximations may be checked for accuracy against 
numerical calculations of the full Einstein equations. See, e.g., [1] 
for a recent review of rotating bodies in GR.

Similarly, in any realistic alternative theory of gravity it must be possible 
to calculate the effects of rotation on gravitational fields. In particular 
this applies to quasi-metric gravity (the quasi-metric framework is described 
in detail elsewhere [2, 3]). In this paper, equations relevant for a metrically
stationary, axially symmetric, isolated system in quasi-metric gravity are set 
up and a first attempt is made to find solutions. However, it is found that
no such solutions exist so that in principle, gravitational fields
corresponding to axially symmetric, spinning sources must be non-stationary.
But for sufficiently weak gravitational fields, approximately stationary 
solutions may still be found.

The paper is organized as follows. Section 2 contains a brief survey of the 
quasi-metric framework. In section 3, the relevant equations are set up and the
gravitational field outside an approximately metrically stationary, axially 
symmetric, isolated spinning source in quasi-metric gravity is calculated 
approximately and compared to the Kerr metric. In the limit of no rotation, one
gets back the spherically symmetric, metrically static case treated in [4] if 
the source cannot support shear forces. In section 4, we calculate the geodetic
effect for a gyroscope in orbit within the quasi-metric framework. Section 5 
contains some concluding remarks.
\section{Quasi-metric gravity described succinctly}  
Quasi-metric relativity (QMR), including its current observational status, is 
described in detail elsewhere [2-4]. Here we give a very brief survey and 
include only the formulae needed for calculations.

The basic idea, which acts as a motivation for postulating the quasi-metric
geometrical framework, is that the cosmic expansion should be described as an 
inherent geometric property of quasi-metric space-time itself and not as a 
kinematical (in the general sense of the word) phenomenon subject to dynamical 
laws. That is, in QMR, the cosmic expansion is described as a general 
phenomenon that does not have a cause. This means that its description should 
not depend on the causal structure associated with any pseudo-Riemannian 
manifold. Such an idea is attractive since in this way, one should be able to 
avoid the in principle enormous multitude of possibilities present, regarding 
cosmic genesis, initial conditions and evolution, if space-time is modelled as
a pseudo-Riemannian manifold. Therefore, one expects that any theory of gravity 
compatible with the quasi-metric framework should be more predictive than any 
metric theory of gravity when it comes to cosmology.

The geometric basis of the quasi-metric framework consists of a 5-dimensional 
differentiable product manifold ${\cal M}{\times}{\bf R}_1$, where 
${\cal M}={\cal S}{\times}{\bf R}_2$ is a (globally hyperbolic) Lorentzian 
space-time manifold, ${\bf R}_1$ and ${\bf R}_2$ are two copies of the real 
line and ${\cal S}$ is a compact Riemannian 3-dimensional manifold 
(without boundaries). The {\em global time function t} is then introduced as a 
coordinate on ${\bf R}_1$. The product topology of ${\cal M}$ implies
that once $t$ is given, there must exist a ``preferred'' ordinary time 
coordinate $x^0$ on ${\bf R}_2$ such that $x^0$ scales like $ct$. A coordinate 
system on ${\cal M}$ with a global time coordinate of this type we call {\em a 
global time coordinate system} (GTCS). Hence, expressed in a GTCS 
${\{}x^{\mu}{\}}$ (where ${\mu}$ can take any value $0-3$), $x^0$ is 
interpreted as a global time coordinate on ${\bf R}_2$ and ${\{}x^j{\}}$ (where
$j$ can take any value $1-3$) as spatial coordinates on ${\cal S}$. The class 
of GTCSs is a set of preferred coordinate systems inasmuch as the equations of 
QMR take special forms in a GTCS. 

The manifold ${\cal M}{\times}{\bf R}_1$ is equipped with two degenerate 
5-dimensional metrics ${\bf {\bar g}}_t$ and ${\bf g}_t$. By definition, the 
``dynamical'' metric ${\bf {\bar g}}_t$ represents a solution of field 
equations, and from ${\bf {\bar g}}_t$ one can construct the ``physical'' 
metric ${\bf g}_t$ which is used when comparing predictions to experiments
involving the equations of motion. To reduce the 5-dimensional space-time 
${\cal M}{\times}{\bf R}_1$ to a 4-dimensional space-time we just slice the 
4-dimensional sub-manifold ${\cal N}$ determined by the equation $x^0=ct$ 
(using a GTCS) out of ${\cal M}{\times}{\bf R}_1$. It is essential that there 
is no arbitrariness in this choice of slicing. That is, the identification of 
$x^0$ with $ct$ must be unique since the two global time coordinates should be 
physically equivalent; the only reason to separate between them is that they 
are designed to parameterize fundamentally different physical phenomena. Note 
that ${\cal S}$ is defined as a compact 3-dimensional manifold to avoid 
ambiguities in the slicing of ${\cal M}{\times}{\bf R}_1$.

Moreover, in ${\cal N}$, ${\bf {\bar g}}_t$ and ${\bf g}_t$ are interpreted as 
one-parameter metric families (this interpretation is merely a matter of 
semantics). Thus by construction, ${\cal N}$ is a 4-dimensional space-time 
manifold equipped with two one-parameter families of Lorentzian 4-metrics 
parameterized by the global time function $t$. This is the general form of the 
quasi-metric space-time framework. We will call ${\cal N}$ 
{\em a quasi-metric space-time manifold.} One reason why ${\cal N}$ 
cannot be represented by single Lorentzian manifolds is that the affine 
connection compatible with any metric family is non-metric; see [2, 3] for 
more details.

From the definition of quasi-metric space-time we see that it is constructed 
as consisting of two mutually orthogonal foliations: on the one hand 
space-time can be sliced up globally into a family of 3-dimensional space-like 
hypersurfaces (called the fundamental hypersurfaces (FHSs)) by the global 
time function $t$, on the other hand space-time can be foliated into a family 
of time-like curves everywhere orthogonal to the FHSs. These curves represent 
the world lines of a family of hypothetical observers called the fundamental 
observers (FOs), and the FHSs together with $t$ represent a preferred notion 
of space and time. That is, the equations of any theory of gravity based on 
quasi-metric geometry should depend on quantities obtained from this preferred
way of splitting up space-time into space and time. 

The metric families ${\bf {\bar g}}_t$ and ${\bf g}_t$ may be decomposed into
parts respectively normal to and intrinsic to the FHSs. The normal parts
involve the unit normal vector field families ${\bf {\bar n}}_t$ and 
${\bf n}_t$ of the FHSs, with the property ${\bf {\bar g}}_t({\bf {\bar n}}_t,
{\bf {\bar n}}_t)={\bf g}_t({\bf n}_t,{\bf n}_t)=-1$. The parts intrinsic to
the FHSs are the spatial metric families ${\bf {\bar h}}_t$ and
${\bf h}_t$, respectively. We may then write
\eqa
{\bf {\bar g}}_t=-{\bf {\bar g}}_t({\bf {\bar n}}_t,{\cdot}){\otimes}
{\bf {\bar g}}_t({\bf {\bar n}}_t,{\cdot})+{\bf {\bar h}}_t,
\ena
and similarly for the decomposition of ${\bf g}_t$. In what follows, we will 
use index notation where Greek indices are ordinary space-time indices taking 
integer values in the range $0..3$, while Latin indices are space indices 
taking integer values in the range $1..3$. Any implicit dependence on $t$ will 
be indicated with a separate index, e.g., the family ${\bf {\bar g}}_t$ has the
space-time coordinates ${\bar g}_{(t){\mu}{\nu}}$. Furthermore, Einstein's 
summation convention will be used throughout. Using said notations and 
expressed in a GTCS, ${\bf {\bar n}}_t$ may be written as
\eqa 
{\bf {\bar n}}_t{\equiv}{\bar n}_{(t)}^{\mu}{\frac{\partial}
{{\partial}x^{\mu}}}={\bar N}_t^{-1}{\Big (}{\frac{\partial}{{\partial}x^0}}-
{\frac{t_0}{t}}{\bar N}^k_{(t)}{\frac{\partial}{{\partial}x^k}}{\Big )},
\ena
where $t_0$ is an arbitrary reference epoch setting the scale of the spatial 
coordinates. Moreover, ${\bar N}_t$ is the family of lapse functions of the FOs
and ${\frac{t_0}{t}}{\bar N}^k_{(t)}$ are the components of the shift vector 
field family of the FOs in $({\cal N},{\bf {\bar g}}_t)$. (A similar formula is
valid for ${\bf n}_t$.) Note that in the rest of this paper we will use the 
symbol `${\bar {\perp}}$' to mean a scalar product with $-{\bf {\bar n}}_t$. A 
useful quantity derived from ${\bar N}_t$ is the 4-acceleration field 
${\bf {\bar a}}_{\cal F}$ of the FOs in $({\cal N},{\bf {\bar g}}_t)$ which is a 
quantity intrinsic to the FHSs. Expressed in a GTCS, ${\bf {\bar a}}_{\cal F}$ 
is defined by its components
\eqa
c^{-2}{\bar a}_{{\cal F}j}{\equiv}{\frac{{\bar N}_t,_j}{{\bar N}_t}}, \qquad 
{\bar y}_t{\equiv}c^{-1}{\sqrt{{\bar a}_{{\cal F}k}{\bar a}_{\cal F}^k}},
\ena
where its norm is given by $c{\bar y}_t$ and where a comma denotes taking a 
partial derivative.

We now write down the most general form allowed for the family 
${\bf {\bar g}}_t$ (expressed in a suitable GTCS) including both explicit and 
implicit dependences on $t$. That is, a general family ${\bf {\bar g}}_t$ can 
be represented by the family of line elements valid on the FHSs (this may be 
taken as a definition)
\eqa
{\overline {ds}}_t^2={\bar N}_t^2{\Big \{ }
[{\bar N}_{(t)}^k{\bar N}_{(t)}^s{\tilde h}_{(t)ks}-1](dx^0)^2+
2{\frac{t}{t_0}}{\bar N}_{(t)}^k{\tilde h}_{(t)ks}dx^sdx^0+
{\frac{t^2}{t_0^2}}{\tilde h}_{(t)ks}dx^kdx^s{\Big \} }.
\ena
Here, $d{\bar{\sigma}}_t^2{\equiv}{\bar h}_{(t)ks}dx^kdx^s{\equiv}
{\frac{t^2}{t_0^2}}{\bar N}_t^2{\tilde h}_{(t)ks}dx^kdx^s$ is the line element 
family corresponding to the spatial metric family ${\bf {\bar h}}_t{\equiv}
{\frac{t^2}{t_0^2}}{\bar N}_t^2{\bf {\tilde h}}_t$ intrinsic to the FHSs.

Field equations from which ${\bar N}_t$ and ${\bar N}^j_{(t)}$ can be determined
are given as couplings between projections of the Ricci tensor family
${\bf {\bar R}}_t$ and the active stress-energy tensor ${\bf T}_t$. Said 
projections are scalar and 3-vector projections respectively, so to have a 
sufficient number of field equations to determine ${\tilde h}_{(t)ks}$, it is 
also necessary to have a 3-tensor field equation. It turns out that this 
equation contains no extra couplings to ${\bf T}_t$. Now it is found from 
dimensional analysis yields that the gravitational coupling in QMR must be 
nonuniversal [2, 3]. That is, there must be two different (variable) 
gravitational coupling parameters $G^{\rm B}_t$ and $G^{\rm S}_t$ for coupling of 
different types of matter sources to space-time geometry. To be more specific, 
$G^{\rm B}_t$ couples to the active electromagnetic stress-energy tensor 
${\bf T}^{\rm (EM)}_t$ while $G^{\rm S}_t$ couples to the active stress-energy 
tensor ${\bf T}^{\rm (MA)}_t$ representing material sources [2, 3]. Two field 
equations, containing independent couplings to projections of ${\bf T}_t$, then 
read (using a GTCS)
\eqa
{\hspace*{-1mm}}2{\bar R}_{(t){\bar {\perp}}{\bar {\perp}}}=
2(c^{-2}{\bar a}_{{\cal F}{\mid}i}^i+
c^{-4}{\bar a}_{{\cal F}i}{\bar a}_{\cal F}^i-{\bar K}_{(t)ik}
{\bar K}_{(t)}^{ik}+{\cal L}_{{\bf {\bar n}}_t}{\bar K}_t) \nonumber \\
={\kappa}^{\rm B}(T^{\rm (EM)}_{(t){\bar {\perp}}{\bar {\perp}}}+
{\hat T}^{{\rm (EM)}i}_{(t)i})+
{\kappa}^{\rm S}(T^{\rm (MA)}_{(t){\bar {\perp}}{\bar {\perp}}}+
{\hat T}^{{\rm (MA)}i}_{(t)i}),
\ena
\eqa
{\bar R}_{(t)j{\bar {\perp}}}+{\Big (}{\frac{{\bar h}_{(t)}^{ik}}{{\bar N}_t}}
{\frac{\partial}{{\partial}x^0}}{\bar h}_{(t)ij}{\Big )}_{{\mid}k}-
{\Big (}{\frac{{\bar h}_{(t)}^{ik}}{{\bar N}_t}}
{\frac{\partial}{{\partial}x^0}}{\bar h}_{(t)ik}{\Big )},_j
={\kappa}^{\rm B}T^{{\rm (EM)}}_{(t)j{\bar {\perp}}}
+{\kappa}^{\rm S}T^{\rm (MA)}_{(t)j{\bar {\perp}}},
\ena
where ${\cal L}_{{\bf {\bar n}}_t}$ denotes a projected Lie derivative (acting on 
spatial objects only) in the direction normal to the FHS and ${\bf {\bar K}}_t$
is the extrinsic curvature tensor family (with trace ${\bar K}_t$) of the FHSs. 
Moreover ${\kappa}^{\rm B}{\equiv}8{\pi}G^{\rm B}/c^4$,
${\kappa}^{\rm S}{\equiv}8{\pi}G^{\rm S}/c^4$, a ``hat'' denotes an object 
projected into the FHSs and the symbol `${\mid}$' denotes taking a spatial 
covariant derivative. The values $G^{\rm B}$ of $G^{\rm B}_t$ and $G^{\rm S}$ of 
$G^{\rm S}_t$, respectively, are by convention chosen as those measured in 
(hypothetical) local gravitational experiments in an empty universe at epoch 
$t_0$. Note that all quantities correspond to the metric family 
${\bf {\bar g}}_t$. 

In addition to field equations (5) and (6), we also have the 3-tensor 
equation [2, 3]
\eqa
{\bar C}_{(t){\bar {\perp}}i{\bar {\perp}j}}={\tilde H}_{(t)ij}
+{\frac{1}{(ct{\bar N}_t)^2}}{\bar h}_{(t)ij},
\ena
involving the Weyl tensor family ${\bf {\bar C}}_t$ in 
$({\cal N},{\bf {\bar g}}_t)$. Equivalently, equation (7) yields
\eqa
{\frac{1}{{\bar N}_t}}{\cal L}_{{\bar N}_t{\bf {\bar n}}_t}{\bar K}_{(t)ij}
+{\bar K}_t{\bar K}_{(t)ij}-{\tilde H}_{(t)ij} \nonumber \\
={\frac{1}{3}}{\Big [}{\bar R}_{(t){\bar {\perp}}{\bar {\perp}}}
-{\bar K}_{(t)ks}{\bar K}_{(t)}^{ks}+{\bar K}_t^2
-c^{-2}{\bar a}_{{\cal F}{\mid}s}^s
-c^{-4}{\bar a}_{{\cal F}}^s{\bar a}_{{\cal F}s}+{\frac{3}{(ct{\bar N}_t)^2}}
{\Big ]}{\bar h}_{(t)ij},
\ena
where ${\tilde H}_{(t)ij}$ are the components of the Einstein tensor family
${\bf {\tilde H}}_t$ obtained from the metric family ${\bf {\tilde h}}_t$.
We notice that in equation (8), the (partial) coupling to ${\bf T}_t$ is via 
the quantity ${\bar R}_{(t){\bar {\perp}}{\bar {\perp}}}$ determined from equation 
(5). Besides, note that taking the trace of equation (8) recovers the (general)
expression for ${\bar R}_{(t){\bar {\perp}}{\bar {\perp}}}$ in equation (5).

It shall be useful to have a coordinate expression for ${\bf {\bar K}}_t$. Such
is given from the well-known (except for the $t$-dependence) formula from 
canonical GR
\eqa
{\bar K}_{(t)ij}={\frac{1}{2{\bar N}_t}}{\Big [}{\frac{t}{t_0}}
({\bar N}_{(t)i{\mid}j}+{\bar N}_{(t)j{\mid}i})-
{\frac{\partial}{{\partial}x^0}}{\bar h}_{(t)ij}{\Big ]},
\ena
\eqa
{\bar K}_t={\frac{t_0}{t}}{\frac{{\bar N}^i_{(t){\mid}i}}{{\bar N}_t}}
-{\frac{1}{2{\bar N}_t}}{\bar h}_{(t)}^{ij}{\frac{\partial}{{\partial}x^0}}
{\bar h}_{(t)ij}.
\ena
We notice that in general, we have that ${\bar R}_{(t)j{\bar {\perp}}}=
{\bar K}_{(t)j{\mid}i}^i-{\bar K}_t,_j$. It is also convenient to have explicit 
expressions for the curvature intrinsic to the FHSs. From equation (4) one 
easily calculates (given a prior-geometric restriction on ${\tilde P}_t$, se 
below)
\eqa
{\bar H}_{(t)ij}=c^{-2}{\Big (}{\bar a}_{{\cal F}{\mid}k}^k- 
{\frac{1}{{\bar N}_t^2t^2}}{\Big )}{\bar h}_{(t)ij}-c^{-4}
{\bar a}_{{\cal F}i}{\bar a}_{{\cal F}j}-c^{-2}
{\bar a}_{{\cal F}i{\mid}j}+{\tilde H}_{(t)ij},
\ena
\eqa
{\bar P}_t={\frac{6}{({\bar N}_tct)^2}}+2c^{-4}{\bar a}_{{\cal F}k}
{\bar a}_{\cal F}^k-4c^{-2}{\bar a}_{{\cal F}{\mid}k}^k, \qquad
{\tilde P}_t={\frac{6}{(ct_0)^2}},
\ena
where ${\bf {\bar H}}_t$ is the Einstein tensor family intrinsic to the FHSs
in $({\cal N},{\bf {\bar g}}_t)$ and ${\bar P}_t$ is the corresponding 
curvature scalar family. Similarly, ${\tilde P}_t$ is the curvature scalar
associated with the metric family ${\bf {\tilde h}}_t$. Note that the form of
${\tilde P}_t$ represents prior 3-geometry of the FHSs and that it does not 
depend explicitly on $t$.

The coordinate expression for the covariant divergence of ${\bf T}_t$ (with $t$
fixed), i.e., ${\bf {\bar {\nabla}}}{\bf {\cdot}}{\bf T}_t$, takes the form
[2, 3]
\eqa
T_{(t){\mu};{\nu}}^{\nu}=
2{\frac{{\bar N}_t,_{\nu}}{{\bar N}_t}}T_{(t){\mu}}^{\nu}
=2c^{-2}{\bar a}_{{\cal F}i}{\hat T}^i_{(t){\mu}}
-2{\frac{{\bar N}_t,_{\bar {\perp}}}{{\bar N}_t}}T_{(t){\bar {\perp}}{\mu}},
\ena
which represents the local conservation laws in QMR. Note that these laws do 
not depend on source composition, and that even the metric approximation of 
equation (13) is different from its counterpart in metric gravity.

To construct ${\bf g}_t$ from ${\bf {\bar g}}_t$, we need the 3-vector field
family ${\bf v}_t$. Expressed in a GTCS, ${\bf v}_t$ by definition has the 
components [2, 3]
\eqa
v^j_{(t)}{\equiv}{\bar y}_t{\bar x}_{\cal F}^j, {\qquad} 
v={\bar y}_t{\sqrt{{\bar h}_{(t)ik}{\bar x}_{\cal F}^i{\bar x}_{\cal F}^k}},
\ena
where $v$ is the norm of ${\bf v}_t$ and ${\bf {\bar x}}_{\cal F}$ is a 3-vector 
field found from the equation
\eqa
{\Big [}{\bar a}_{{\cal F}{\mid}k}^k+c^{-2}{\bar a}_{{\cal F}k}
{\bar a}_{\cal F}^k{\Big ]}{\bar x}_{\cal F}^j-
{\Big [}{\bar a}_{{\cal F}{\mid}k}^j+c^{-2}{\bar a}_{{\cal F}k}
{\bar a}_{\cal F}^j{\Big ]}{\bar x}_{\cal F}^k-2{\bar a}_{\cal F}^j=0.
\ena
We now define the unit vector field 
${\bf {\bar e}}_{\cal F}{\equiv}{\frac{t_0}{t}}{\bar e}_{\cal F}^i
{\frac{\partial}{{\partial}x^i}}$ and the corresponding covector field
${\bf {\bar {\omega}}}_{\cal F}{\equiv}{\frac{t}{t_0}}
{\bar {\omega}}_{{\cal F}i}dx^i$ along ${\bf {\bar x}}_{\cal F}$. Then we have 
[2, 3]
\eqa
g_{(t)00}={\Big (}1-{\frac{v^2}{c^2}}{\Big )}^2{\bar g}_{(t)00},
\ena
\eqa
g_{(t)0j}={\Big (}1-{\frac{v^2}{c^2}}{\Big )}{\Big [}{\bar g}_{(t)0j}+
{\frac{t}{t_0}}{\frac{2{\frac{v}{c}}}{1-{\frac{v}{c}}}}
({\bar e}^i_{\cal F}{\bar N}_{(t)i})
{\bar {\omega}}_{{\cal F}j}{\Big ]},
\ena
\eqa
g_{(t)ij}={\bar g}_{(t)ij}+{\frac{t^2}{t_0^2}}
{\frac{4{\frac{v}{c}}}{(1-{\frac{v}{c}})^2}}
{\bar {\omega}}_{{\cal F}i}{\bar {\omega}}_{{\cal F}j}.
\ena
These formulae define the transformation ${\bf {\bar g}}_t{\rightarrow}
{\bf g}_t$. Note that we have eliminated any possible $t$-dependence of 
${\bar N}_t$ in equations (16)-(18) by setting $t=x^0/c$ where it occurs. This 
implies that $N$ does not depend explicitly on $t$.
\section{Metrically stationary, axially symmetric systems}
In this section, we examine the gravitational field interior respectively 
exterior to an isolated, axially symmetric, spinning source made of perfect 
fluid. We also assume that the system has no net translatory motion with 
respect to the cosmic rest frame; i.e., the frame which with respect to the 
FOs are taken to be at rest, on average. (See refs. [2-4] for a discussion of 
why this assumption is not crucial.)

We require that the rotation of the source should have no time 
dependence apart from the effects coming from the global cosmic expansion. 
Besides, as for the spherically symmetric, metrically static case treated in 
[4], we require that the only time dependence of the gravitational field is 
via the cosmic scale factor. But contrary to the metrically static case, there 
is a non-zero shift vector field present due to the rotation of the source. 
However, we require that a GTCS can be found where 
${\bar N}_t$, ${\bar N}_{(t)j}$ and ${\tilde h}_{(t)ij}$ are all independent of 
$x^0$ and $t$. We call this a {\em metrically stationary} case.

The axial symmetry can be directly imposed on equation (4). Introducing a 
spherical GTCS ${\{}x^0,{\rho},{\theta},{\phi}{\}}$ where ${\rho}$ is an 
isotropic radial coordinate and where the shift vector field points in the 
negative ${\phi}$-direction, ${\bar N}_t$ and ${\bar N}_{(t){\phi}}$ do not 
depend on ${\phi}$. Neither do the components of ${\bf {\tilde h}}_t$,
and equation (4) takes the sufficiently general form
\eqa
{\overline{ds}}^2_t={\Big [}{\bar N}_{(t){\phi}}{\bar N}^{\phi}_{(t)}
-{\bar N}_t^2{\Big ]}(dx^0)^2+2{\frac{t}{t_0}}{\bar N}_{(t){\phi}}d{\phi}dx^0
\nonumber \\
+{\frac{t^2}{t^2_0}}{\bar N}^2_t{\Big [}
{\frac{{\bar X}d{\rho}^2}{1-{\frac{{\rho}^2}{{\Xi}_0^2}}}}
+{\bar Y}{\rho}^2d{\theta}^2+{\bar Z}{\rho}^2{\sin}^2{\theta}d{\phi}^2{\Big ]},
\ena
where ${\bar X}$, ${\bar Y}$ and ${\bar Z}$ are functions to be determined,
${\Xi}_0{\equiv}ct_0$ and $0{\leq}{\bar N}_t{\rho}<{\Xi}_0$. The range 
of ${\rho}$ is limited for both physical and mathematical reasons; truly 
isolated systems cannot exist in quasi-metric gravity [4]. That is, realistic
nontrivial global solutions of the field equations on the FHSs for isolated
systems do not exist, according to the maximum principle applied to closed
Riemannian manifolds. However, for metrically static, isolated systems 
${\tilde h}_{(t)ks}=S_{ks}$, where $S_{ks}dx^kdx^s$ is the metric of the 3-sphere 
${\bf S}^3$ (with radius equal to $ct_0$). For this case, QMR allows exact 
``semiglobal'' vacuum solutions on {\em half of} ${\bf S}^3$ (with the 
reasonable boundary condition ${\bar N}_t({\Xi}_0)=1$; the same values as for 
an empty universe). And although such solutions may be mathematically extended 
to (almost) the whole of ${\bf S}^3$, the metric transformation 
${\bf {\bar g}}_t{\rightarrow}{\bf g}_t$ is singular at the boundary, so that 
any transformation based on the extended solutions becomes mathematically 
meaningless. The requirement that both ${\bf {\bar g}}_t$ and ${\bf g}_t$ exist
is thus satisfied only for the half of ${\bf S}^3$ for metrically static, 
vacuum solutions exterior to a metrically static source. If semiglobal 
solutions also exist for metrically stationary vacua remains to be seen.
 
Using the definition ${\bar B}{\equiv}{\bar N}^2_t$, equation (19)
may conveniently be rewritten in the form
\eqa
{\overline{ds}}^2_t={\bar B}{\Big [}-(1-{\bar V}^2{\bar Z}{\rho}^2
{\sin}^2{\theta})(dx^0)^2
+2{\frac{t}{t_0}}{\bar V}{\bar Z}{\rho}^2{\sin}^2{\theta}d{\phi}dx^0
\nonumber \\
+{\frac{t^2}{t^2_0}}{\Big (}{\frac{{\bar X}d{\rho}^2}{1-{\frac{{\rho}^2}
{{\Xi}_0^2}}}}+{\bar Y}{\rho}^2d{\theta}^2+{\bar Z}{\rho}^2{\sin}^2{\theta}
d{\phi}^2{\Big )}{\Big ]},
\ena
where 
\eqa
{\bar V}{\equiv}
{\frac{{\bar N}_{(t){\phi}}}{{\bar B}{\bar Z}{\rho}^2{\sin}^2{\theta}}}.
\ena
Simple expressions for the the non-vanishing components of the extrinsic 
curvature tensor family may be found from equations (9), (20) and (21). They 
read (note that the trace ${\bar K}_t$ vanishes)
\eqa
{\bar K}_{(t){\rho}{\phi}}={\bar K}_{(t){\phi}{\rho}}={\frac{t}{2t_0}}
{\sqrt{\bar B}}{\bar Z}{\rho}^2{\sin}^2{\theta}{\bar V},_{\rho}, \qquad
{\bar K}_{(t){\theta}{\phi}}={\bar K}_{(t){\phi}{\theta}}={\frac{t}{2t_0}}
{\sqrt{\bar B}}{\bar Z}{\rho}^2{\sin}^2{\theta}{\bar V},_{\theta}.
\ena
The unknown quantities ${\bar B}({\rho},{\theta})$, 
${\bar V}({\rho},{\theta})$, ${\bar X}({\rho},{\theta})$, 
${\bar Y}({\rho},{\theta})$ and ${\bar Z}({\rho},{\theta})$ may now in 
principle be calculated from the field equations and local conservation laws.
\subsection{The interior field}
We will now set up the general field equations interior to the source, which 
is modelled as a perfect fluid (in general consisting of a mixture of photons 
and material particles). However, no attempt will be made to find solutions.
Note that the metrically stationary condition implies a negligible net rate of 
energy transfer between photons and material particles. To begin with we 
consider ${\bf T}_t$ for a perfect fluid,
\eqa
{\bf T}_t=({\tilde {\varrho}}_{\rm m}+
c^{-2}{\tilde {p}}){\bf {\bar u}}_t{\otimes}
{\bf {\bar u}}_t + {\tilde p}{\bf {\bar g}}_t, 
\ena
where ${\tilde {\varrho}}_{\rm m}$ is the active mass-energy density in the 
local rest frame of the fluid and ${\tilde p}$ is the active pressure.
Furthermore ${\bf {\bar u}}_t$ is the 4-velocity vector family in 
$({\cal N},{\bf {\bar g}}_t)$ of observers co-moving with the fluid. It is
useful to set up the general formula for the split-up of ${\bf {\bar u}}_t$ 
into pieces respectively normal to and intrinsic to the FHSs:
\eqa
{\bf {\bar u}}_t= {\topstar{\bar {\gamma}}}(c{\bf {\bar n}}_t+
{\bf {\bar w}}_t), \qquad {\topstar{\bar {\gamma}}}{\equiv}
(1-{\frac{{\bar w}^2}{c^2}})^{-{\frac{1}{2}}},
\ena
where ${\bf {\bar w}}_t$ (with norm ${\bar w}$) is the 3-velocity family with 
respect to the FOs. Note that due to the axial symmetry, ${\bf {\bar w}}_t$
points in the ${\pm}{\phi}$-direction. Moreover, by definition the quantity 
${\varrho}_{\rm m}$ is the passive mass-energy density as measured in the 
local rest frame of the fluid and $p$ is the passive pressure. The relationship
between ${\tilde {\varrho}}_{\rm m}$ and ${\varrho}_{\rm m}$ is given by
\eqa
{\varrho}_{\rm m}=
\left\{
\begin{array}{ll}
{\frac{t_0}{t}}{\bar N}_t^{-1}{\tilde {\varrho}}_{\rm m}, 
& \qquad $$\hbox{for a fluid of material particles,}$$ \\[1.5ex]
{\frac{t_0^2}{t^2}}{\bar N}_t^{-2}{\tilde {\varrho}}_{\rm m}, 
& \qquad $$\hbox{for the electromagnetic field,}$$
\end{array}
\right.
\ena
and a similar relationship exists between ${\tilde p}$ and $p$. The reason why
the relationship between ${\tilde {\varrho}}_{\rm m}$ and ${\varrho}_{\rm m}$
is different for the electromagnetic field than for material perfect fluid 
sources, is that the two different gravitational coupling parameters for these
two sources depend differently on $t$ and ${\bar N}_t$ [2, 3], and that by 
convention, these dependences have been transferred to the active mass-energy 
densities.

The next step is to use equations (23) and (24) to find suitable expressions
for the source terms of the field equations (5) and (6). We find
\eqa
T_{(t){\bar {\perp}}{\bar {\perp}}}+{\hat T}_{(t)i}^i=
{\topstar{\bar {\gamma}}}^2(1+{\frac{{\bar w}^2}{c^2}})
({\tilde {\varrho}}_{\text m}c^2+{\tilde p})+2{\tilde p}{\equiv}
{\frac{t_0^2}{t^2{\bar B}}}{\Big [}
{\topstar{\bar {\gamma}}}^2(1+{\frac{{\bar w}^2}{c^2}})
({\bar {\varrho}}_{\text m}c^2+{\bar p})+2{\bar p}{\Big ]},
\ena
where ${\bar {\varrho}}_{\text m}$ is the properly scaled density of active
mass and ${\bar p}$ is the associated pressure. Moreover, we find (assuming
that the source rotates in the positive ${\phi}$-direction)
\eqa
T_{(t){\bar {\perp}}{\phi}}=
{\topstar{\bar {\gamma}}}^2{\frac{{\bar w}_{(t){\phi}}}{c}}
({\tilde {\varrho}}_{\text m}c^2+{\tilde p})
={\frac{t_0}{t}}{\sqrt{\frac{\bar Z}{\bar B}}}{\topstar{\bar {\gamma}}}^2
{\rho}{\sin}{\theta}{\frac{{\bar w}}{c}}
({\bar {\varrho}}_{\text m}c^2+{\bar p}).
\ena
The nontrivial parts of the local conservation laws (12) yield 
\eqa
{\bar p},_{\rho}&=&-{\Big [}{\bar {\varrho}}_{\text m}c^2-3{\bar p}{\Big ]}
{\frac{{\bar B},_{\rho}}{2{\bar B}}}+{\topstar {\bar {\gamma}}}^2
{\frac{{\bar w}}{c}}({\bar {\varrho}}_{\text m}c^2+{\bar p}){\Big [}
{\sqrt{\bar Z}}{\rho}{\sin}{\theta}{\bar V},_{\rho}+{\frac{{\bar w}}{c}}
({\frac{1}{\rho}}+{\frac{{\bar Z},_{\rho}}{2{\bar Z}}}){\Big ]},
\nonumber \\
{\bar p},_{\theta}&=&-{\Big [}{\bar {\varrho}}_{\text m}c^2-3{\bar p}{\Big ]}
{\frac{{\bar B},_{\theta}}{2{\bar B}}}+{\topstar {\bar {\gamma}}}^2
{\frac{{\bar w}}{c}}({\bar {\varrho}}_{\text m}c^2+{\bar p}){\Big [}
{\sqrt{\bar Z}}{\rho}{\sin}{\theta}{\bar V},_{\theta}+{\frac{{\bar w}}{c}}
({\cot}{\theta}+{\frac{{\bar Z},_{\theta}}{2{\bar Z}}}){\Big ]}.
\ena
Note that the metrically stationary condition implies that one must have an 
equation of state of the form $p{\propto}{\varrho}_{\text m}$ since otherwise
${\bar {\varrho}}_{\text m}$ and ${\bar p}$ cannot both be independent of $t$.

We are now in position to set up the field equations (5), (6) for the system 
in an as simple as possible form. (However, the extra field equation (8) turns
out to yield some very complicated expressions that will be more suitably
presented in an appendix.) After calculating the necessary derivatives and 
doing some simple algebra we find the two coupled partial differential 
equations (with ${\bar p}^{\rm (EM)}={\frac{1}{3}}{\bar {\varrho}}^{\rm (EM)}c^2$)
\eqa
{\bar Y}(1-{\frac{{\rho}^2}{{\Xi}_0^2}}){\bar B},_{{\rho}{\rho}}
+{\frac{{\bar X}}{{\rho}^2}}{\bar B},_{{\theta}{\theta}}
+{\bar Y}{\Big [}{\frac{2}{\rho}}(1-{\frac{3{\rho}^2}{2{\Xi}_0^2}})
+{\frac{1}{2}}(1-{\frac{{\rho}^2}{{\Xi}_0^2}})
({\frac{{\bar Y},_{\rho}}{\bar Y}}-{\frac{{\bar X},_{\rho}}{\bar X}}
+{\frac{{\bar Z},_{\rho}}{\bar Z}}){\Big ]}{\bar B},_{\rho} \nonumber \\
+{\frac{{\bar X}}{{\rho}^2}}{\Big [}{\cot}{\theta}+
{\frac{1}{2}}({\frac{{\bar X},_{\theta}}{\bar X}}-
{\frac{{\bar Y},_{\theta}}{\bar Y}}+{\frac{{\bar Z},_{\theta}}{\bar Z}})
{\Big ]}{\bar B},_{\theta} \nonumber \\
={\bar B}{\Big {\{}}{\bar Z}{\rho}^2{\sin}^2{\theta}{\Big [}{\bar Y}
(1-{\frac{{\rho}^2}{{\Xi}_0^2}})({\bar V},_{\rho})^2+
{\frac{{\bar X}}{{\rho}^2}}({\bar V},_{\theta})^2{\Big ]}
+{\frac{2}{3}}{\bar X}{\bar Y}{\kappa}^{\rm B}
{\bar {\varrho}}^{\rm (EM)}_{\text m}c^2{\Big [}
2{\topstar {\bar {\gamma}}}^2(1+{\frac{{\bar w}^2}{c^2}})+1{\Big ]} 
\nonumber \\
+{\bar X}{\bar Y}{\kappa}^{\rm S}{\Big [}{\topstar {\bar {\gamma}}}^2
(1+{\frac{{\bar w}^2}{c^2}})({\bar {\varrho}}^{\rm (MA)}_{\text m}c^2+
{\bar p}^{\rm (MA)})+2{\bar p}^{\rm (MA)}{\Big ]}{\Big {\}}},
\ena
\eqa
{\bar Y}(1-{\frac{{\rho}^2}{{\Xi}_0^2}}){\bar V},_{{\rho}{\rho}}+
{\frac{{\bar X}}{{\rho}^2}}{\bar V},_{{\theta}{\theta}}+{\bar Y}
{\Big [}{\frac{4}{\rho}}-{\frac{5{\rho}}{{\Xi}_0^2}}+
(1-{\frac{{\rho}^2}{{\Xi}_0^2}})({\frac{{\bar B},_{\rho}}{\bar B}}
+{\frac{3{\bar Z},_{\rho}}{2{\bar Z}}}-{\frac{{\bar X},_{\rho}}{2{\bar X}}}
+{\frac{{\bar Y},_{\rho}}{2{\bar Y}}}){\Big ]}{\bar V},_{\rho} \nonumber \\
+{\frac{{\bar X}}{{\rho}^2}}{\Big [}3{\cot}{\theta}+
{\frac{{\bar B},_{\theta}}{\bar B}}+{\frac{3{\bar Z},_{\theta}}{2{\bar Z}}}-
{\frac{{\bar Y},_{\theta}}{2{\bar Y}}}+{\frac{{\bar X},_{\theta}}{2{\bar X}}}
{\Big ]}{\bar V},_{\theta} \nonumber \\
={\bar X}{\bar Y}{\Big [}{\frac{8}{3}}{\kappa}^{\rm B}
{\topstar{\bar {\gamma}}}^2{\frac{{\bar w}}{c}}
{\frac{{\bar {\varrho}}^{\rm (EM)}_{\text m}c^2}
{{\rho}{\sin}{\theta}}}+2{\kappa}^{\rm S}{\topstar{\bar {\gamma}}}^2
{\frac{{\bar w}}{c}}{\frac{({\bar {\varrho}}^{\rm (MA)}_{\text m}c^2+
{\bar p}^{\rm (MA)})}{{\rho}{\sin}{\theta}}}{\Big ]}.
\ena
Note that for the special case when ${\bar X}={\bar Y}={\bar Z}=1$,
the left hand side of equation (29) is equal to ${\nabla}_{\text s}^2{\bar B}$, 
where ${\nabla}_{\text s}^2$ is the Laplacian compatible with the standard metric
on ${\bf S}^3$ (with radius equal to ${\Xi}_0$).

To avoid problems with coordinate pathologies along the axis of rotation, it 
would probably be convenient to express equations (28)-(30) in Cartesian
coordinates rather than trying to solve them numerically as they stand (in 
combination with equation (8)). However, whenever ${\bar K}_t{\equiv}0$, 
equation (8) for the interior field takes the same form as for the exterior 
field (see the appendix). Moreover, it can be shown that equation (8) has no 
solution for the axisymmetric, metrically stationary case (see the next 
section).
\subsection{The exterior field}
For illustrative purposes, let us first consider exact solutions of equation 
(29) without source terms for the {\em metrically static}, axisymmetric case; 
we may then set ${\bar V}=0$ and ${\bar X}={\bar Y}={\bar Z}=1$. If we also 
insist that the solution ${\bar B}^{{\rm {(ms)}}}({\rho},{\theta})$ fulfils the 
boundary condition ${\bar B}^{{\rm {(ms)}}}({\Xi}_0,{\theta})=1$ (which should not
be taken to be a realistic {\em physical} constraint, since true isolated 
systems do not exist in QMR [4]), it turns out that the solution (on the half 
of ${\bf S}^3$) must take the form
\eqa
{\bar B}^{{\rm {(ms)}}}({\rho},{\theta})=1-{\frac{r_{\rm s0}}{\rho}}
{\sqrt{1-{\frac{{\rho}^2}{{\Xi}_0^2}}}}{\Big [}1-J_{\rm 2}{\frac
{{\bar {\cal R}}^2}{2{\rho}^2}}(3{\cos}^2{\theta}-1){\Big ]},
\ena
where $J_{\rm 2}$ is the (static) quadrupole-moment parameter and the other
quantities are as in equation (32) below. It is clear that the source 
corresponding to this solution is a body which has a nonspherical shape 
(oblate spheroid) in absence of any rotation. Thus this body cannot be made of 
perfect fluid, since the source material must be able to support shear forces. 
From the solution (31) one is in principle able to construct the counterpart 
exact ``physical'' metric family ${\bf g}_t^{\rm {(ms)}}$ using equations 
(14)-(18). However, the expressions thus obtained are extremely complicated so 
we will not include them here. A series expansion can be obtained from 
equation (41) below in the limit of no rotation with a non-zero static 
quadrupole-moment parameter. Note that the solution (31) can be extended to 
(almost) the whole of ${\bf S}^3$ but that the associated family 
${\bf g}^{\rm {(ms)}}_t$ cannot.

Returning to the metrically stationary, axisymmetric case; before one tries
to find an exact exterior solution of equations (29), (30) (without sources) 
in combination with equations (A.2)-(A.5), it would seem reasonable first 
trying to find the first few terms of a candidate trial series solution. 
However, when one tries to solve the equation set (A.2)-(A.5) with the lowest 
order expression for ${\bar V}$ inserted, one finds that it has no solution
(see the appendix). The trivial solution ${\bar X}={\bar Y}={\bar Z}=1$ works 
up to $O(6)$ in small quantities, but for $O(8)$ the lowest order contribution 
from ${\bar V}$ shows up, and it is found that no trial solution exists to this 
order. (Note that ${\frac{r_{\text s0}}{\rho}}$, e.g., is of $O(2)$, see below.)

So we are forced to conclude that {\em in quasi-metric gravity, no axially
symmetric, metrically stationary systems exist}. This means that in
quasi-metric gravity, axisymmetric systems associated with rotating sources 
must necessarily be non-stationary. However, for weak fields and slow 
rotations, such systems may still be considered as stationary to a very good 
approximation. Therefore, it is still meaningful to find an approximate series 
solution of equations (29) and (30) up to an order that will not be 
inconsistent with the trivial approximate solution 
${\bar X}={\bar Y}={\bar Z}=1$ of equations (A.2)-(A.5). Since this approximate
solution works up to $O(6)$, we see from equation (29) and (30), respectively, 
that ${\bar B}$ may be considered stationary up to $O(6)$ and 
${\rho}{\bar V}$ may be considered stationary up to $O(5)$.

It is straightforward to calculate the first few terms of the approximate 
series solution (for the case where the source spins in the positive 
${\phi}$-direction). We find
\eqa
{\bar B}({\rho},{\theta})=1-{\frac{r_{\text s0}}{\rho}}+
{\frac{r_{\text s0}{\rho}}{2{\Xi}_0^2}}+
J_2{\frac{{\bar {\cal R}}^2r_{\text s0}}{2{\rho}^3}}(3{\cos}^2{\theta}-1)+
{\frac{r_{\text s0}^2a_0^2}{2{\rho}^4}}(3{\sin}^2{\theta}-1) \nonumber \\
+{\rm {\ }higher{\ }order{\ } nonstationary{\ }terms},
\ena
\eqa
{\bar V}({\rho},{\theta})=-{\frac{r_{\text s0}a_0}{{\rho}^3}}{\Big (}1+
{\frac{3r_{\text s0}}{4{\rho}}}+{\cdots}{\Big )},
\ena
where $J_2$ now is the rotationally induced quadrupole-moment parameter and
${\bar {\cal R}}$ is the mean coordinate radius of the source. Furthermore,
$r_{\text s0}{\equiv}2M^{\rm (EM)}_{t_0}G^{\rm B}/c^2+2M^{\rm (MA)}_{t_0}G^{\rm S}/c^2$ is 
the generalized Schwarzschild radius at epoch $t_0$ defined from the Komar 
masses [5, 6] $M^{\rm (EM)}_{t_0}$ and $M^{\rm (MA)}_{t_0}$, i.e.
\eqa
M^{\rm (MA)}_{t_0}{\equiv}c^{-2}{\int}{\int}{\int}{\Big [}{\bar N}_{t_0}
(T^{\rm (MA)}_{(t_0){\bar {\perp}}{\bar {\perp}}}+
{\hat T}^{{\rm (MA)}i}_{(t_0)i})-
2{\bar N}^{\phi}_{(t_0)}T^{\rm (MA)}_{(t_0){\bar {\perp}}{\phi}}{\Big ]}
{\sqrt{{\bar h}_{t_0}}}d^3x,
\ena
and a similar formula for $M^{\rm (EM)}_{t_0}$. Moreover, $a_0$ is a length at 
epoch $t_0$ defined from the angular momentum integrals [6] $J^{\rm (EM)}_{t_0}$ 
and $J^{\rm (MA)}_{t_0}$ (the integrals are in principle taken over the half of 
${\bf S}^3$ since, as for the solution (31), one expects that any further 
extension of the series solution would be inconsistent and mathematically 
irrelevant)
\eqa
a_0{\equiv}{\frac{J^{\rm (EM)}_{t_0}G^{\rm B}+J^{\rm (MA)}_{t_0}G^{\rm S}}
{c(M^{\rm (EM)}_{t_0}G^{\rm B}+M^{\rm (MA)}_{t_0}G^{\rm S})}}, \qquad
J^{\rm (MA)}_{t_0}{\equiv}c^{-1}{\int}{\int}{\int}{\psi}^{\phi}
T^{\rm (MA)}_{(t_0){\bar {\perp}}{\phi}}{\sqrt{{\bar h}_{t_0}}}d^3x,
\ena
and a similar formula for $J^{\rm (EM)}_{t_0}$. Here, ${\bar h}_{t_0}$ is the 
determinant of ${\bf {\bar h}}_{t_0}$ and 
${\bf {\psi}}{\equiv}{\frac{\partial}{{\partial}{\phi}}}$ is a Killing
vector field associated with the axial symmetry. Note that 
$J_t=J^{\rm (EM)}_t+J^{\rm (MA)}_t={\frac{t^2}{t_0^2}}J_{t_0}$ is the active angular 
momentum of the source at epoch $t$. (The corresponding passive angular 
momentum for a purely material source (i.e., containing no photons) is 
$L^{\rm (MA)}_t={\frac{t}{t_0}}L^{\rm (MA)}_{t_0}$, where 
$L^{\rm (MA)}_{t_0}=c^{-1}{\int}{\int}{\int}{\psi}^{\phi}
{\bar {\cal T}}^{\rm (MA)}_{(t_0){\bar {\perp}}{\phi}}{\sqrt{{\bar h}_{t_0}}}d^3x$, and 
where ${\bar {\cal T}}^{\rm (MA)}_t$ is the passive stress-energy tensor for a
purely material source in $({\cal N},{\bf {\bar g}}_t)$.) We now insert 
equations (32) and (33) into equation (20). Taking into account a relevant 
number of terms we get
\eqa
{\overline{ds_t}}^2=-{\Big (}1-{\frac{r_{\rm s0}}{\rho}}+
{\frac{r_{\text s0}{\rho}}{2{\Xi}_0^2}}+
J_2{\frac{{\bar {\cal R}}^2r_{\text s0}}{2{\rho}^3}}(3{\cos}^2{\theta}-1)-
{\frac{r_{\text s0}^2a_0^2}{2{\rho}^4}}{\cos}^2{\theta}+{\cdots}{\Big )}
(dx^0)^2 \nonumber \\
-2{\frac{t}{t_0}}(1-{\frac{r_{\text s0}}{4{\rho}}}+{\cdots})
{\frac{r_{\text s0}a_0}{{\rho}}}{\sin}^2{\theta}d{\phi}dx^0+{\frac{t^2}{t^2_0}}
(1-{\frac{r_{\text s0}}{\rho}}+{\cdots}){\Big (}
{\frac{d{\rho}^2}{1-{\frac{{\rho}^2}{{\Xi}_0^2}}}}+{\rho}^2d{\Omega}^2{\Big )}.
\ena
To construct ${\bf g}_t$ from ${\bf {\bar g}}_t$ we need to calculate the 
vector field ${\bf {\bar x}}_{\cal F}$ from equation (15) 
(${\bf {\bar a}}_{\cal F}$ and its derivatives may be found from equations (3) 
and (32)). These calculations get quite complicated so it is convenient to do 
them by computer. The result is
\eqa
{\bar x}_{\cal F}^{\rho}={\rho}{\Big (}1-{\frac{r_{\rm s0}}{2{\rho}}}-
{\frac{r_{\rm s0}^2}{4{\rho}^2}}+
J_2{\frac{3{\bar {\cal R}}^2}{2{\rho}^2}}(3{\cos}^2{\theta}-1)
-{\frac{r_{\text s0}{\rho}}{4{\Xi}_0^2}}-{\frac{r_{\rm s0}^3}{8{\rho}^3}} 
\nonumber \\ +{\frac{3r_{\text s0}a_0^2}{2{\rho}^3}}(3{\sin}^2{\theta}-2)-
J_2{\frac{3{\bar {\cal R}}^2r_{\text s0}}{4{\rho}^3}}(3{\cos}^2{\theta}-1)+
{\cdots}{\Big )},
\ena
\eqa
{\rho}{\bar x}_{\cal F}^{\theta}=-{\frac{3{\sin}(2{\theta})}{{\rho}}}{\Big (}
J_2{\bar {\cal R}}^2(1-{\frac{3r_{\rm s0}}{4{\rho}}}+{\cdots})-
{\frac{3r_{\text s0}a_0^2}{2{\rho}}}+{\cdots}{\Big )}.
\ena
Furthermore, we need the quantity $v$ defined in equation (14). This may be
expressed by ${\bar B}$ and its derivatives together with the components of
${\bf {\bar x}}_{\cal F}$. We find
\eqa
v={\frac{c}{2{\bar B}}}{\sqrt{(1-{\frac{{\rho}^2}{{\Xi}_0^2}})
({\bar B},_{\rho})^2+{\rho}^{-2}({\bar B},_{\theta})^2}}{\sqrt{
(1-{\frac{{\rho}^2}{{\Xi}_0^2}})^{-1}
({\bar x}_{\cal F}^{\rho})^2+({\rho}{\bar x}_{\cal F}^{\theta})^2}}
={\frac{r_{\rm s0}c}{2{\rho}}}{\Big (}1+{\frac{r_{\rm s0}}{2{\rho}}}
\nonumber \\ +{\frac{r_{\rm s0}^2}{4{\rho}^2}}+{\frac{{\rho}^2}{2{\Xi}_0^2}}
+{\frac{r_{\rm s0}^3}{8{\rho}^3}}-{\frac{r_{\rm s0}{\rho}}{2{\Xi}_0^2}}-
J_2{\frac{{\bar {\cal R}}^2r_{\text s0}}{2{\rho}^3}}(3{\cos}^2{\theta}-1)-
{\frac{r_{\text s0}a_0^2}{2{\rho}^3}}(3{\sin}^2{\theta}+2)+{\cdots}{\Big )}. 
\ena
Finally, to do the transformation shown in equation (18), we need the 
quantities ${\bar {\omega}}_{{\cal F}{\rho}}$ and 
${\bar {\omega}}_{{\cal F}{\theta}}$. Since ${\bar {\omega}}_{{\cal F}{\theta}}$ is 
equal to ${\rho}{\bar x}_{\cal F}^{\theta}$ to the accuracy calculated here, it is
sufficient to write down the expression for ${\bar {\omega}}_{{\cal F}{\rho}}$. A 
straightforward calculation yields
\eqa
{\bar {\omega}}_{{\cal F}{\rho}}=1-{\frac{r_{\rm s0}}{2{\rho}}}-
{\frac{r_{\rm s0}^2}{8{\rho}^2}}+{\frac{{\rho}^2}{2{\Xi}_0^2}}
+{\frac{r_{\text s0}{\rho}}{4{\Xi}_0^2}}-{\frac{r_{\rm s0}^3}{16{\rho}^3}}+
J_2{\frac{{\bar {\cal R}}^2r_{\text s0}}{4{\rho}^3}}(3{\cos}^2{\theta}-1)+
{\cdots},
\ena
and the transformations (16)-(18) then yield, to desired accuracy
\eqa
{ds_t}^2=-{\Big (}1-{\frac{r_{\rm s0}}{\rho}}-
{\frac{r_{\text s0}^2}{2{\rho}^2}}+{\frac{r_{\text s0}{\rho}}{2{\Xi}_0^2}}+
J_2{\frac{{\bar {\cal R}}^2r_{\text s0}}{2{\rho}^3}}(3{\cos}^2{\theta}-1)
-{\frac{r_{\text s0}^2a_0^2}{2{\rho}^4}}{\cos}^2{\theta} \nonumber \\
+{\frac{3r_{\text s0}^4}{16{\rho}^4}}-{\frac{r_{\text s0}^2}{2{\Xi}_0^2}}+
{\cdots}{\Big )}(dx^0)^2-2{\frac{t}{t_0}}(1-{\frac{r_{\text s0}}{4{\rho}}}+
{\cdots}){\frac{r_{\text s0}a_0}{{\rho}}}{\sin}^2{\theta}d{\phi}dx^0
\nonumber \\ +{\frac{t^2}{t_0^2}}
{\Big (}(1+{\frac{r_{\rm s0}}{\rho}}+{\frac{r_{\text s0}^2}{{\rho}^2}}+
{\cdots}){\frac{d{\rho}^2}{1-{\frac{{\rho}^2}{{\Xi}_0^2}}}}+
(1-{\frac{r_{\text s0}}{{\rho}}}+O({\frac{r_{\rm s0}^3}{{\rho}^3}})){\rho}^2
d{\Omega}^2{\Big )}.
\ena
Note that to $O({\frac{r_{\rm s0}^3}{{\rho}^3}})=O(6)$ or higher, the spatial
metric family ${\bf h}_t$ is not diagonal in these coordinates. It will also 
be convenient to express the line element family (41) in an ``almost
Schwarzschild'' radial coordinate $r$ defined by
\eqa
r{\equiv}{\rho}{\sqrt{1-{\frac{r_{\rm s0}}{\rho}}
{\sqrt{1-{\frac{{\rho}^2}{{\Xi}_0^2}}}}}}&=&
{\Big (}1-{\frac{r_{\text s0}}{2{\rho}}}-{\frac{r^2_{\text s0}}{8{\rho}^2}}
-{\frac{r^3_{\text s0}}{16{\rho}^3}}+{\frac{r_{\text s0}{\rho}}{4{\Xi}_0^2}}
+{\cdots}{\Big )}{\rho}, \nonumber \\
{\rho}&=&{\Big (}1+{\frac{r_{\text s0}}{2r}}+{\frac{r^2_{\text s0}}{8r^2}}
-{\frac{r_{\text s0}r}{4{\Xi}_0^2}}+{\cdots}{\Big )}r.
\ena
That is, to the order in small quantities considered in equation (41), at
epoch $t_0$ the surface area of spheres centered on the origin is equal to 
$4{\pi}r^2$. Expressed in the new radial coordinate, equation (41) reads
\eqa
{ds_t}^2=-{\Big (}1-{\frac{r_{\rm s0}}{r}}+
{\frac{3r_{\text s0}^3}{8r^3}}+{\frac{r_{\text s0}r}{2{\Xi}_0^2}}+
J_2{\frac{{\bar {\cal R}}^2r_{\text s0}}{2r^3}}(3{\cos}^2{\theta}-1)(1-
{\frac{3r_{\text s0}}{2r}})
-{\frac{r_{\text s0}^2a_0^2}{2r^4}}{\cos}^2{\theta} \nonumber \\
-{\frac{r_{\text s0}^4}{16r^4}}-{\frac{r_{\text s0}^2}{2{\Xi}_0^2}}+
{\cdots}{\Big )}(dx^0)^2-2{\frac{t}{t_0}}(1-{\frac{3r_{\text s0}}{4r}}+
{\cdots}){\frac{r_{\text s0}a_0}{r}}{\sin}^2{\theta}d{\phi}dx^0
\nonumber \\ +{\frac{t^2}{t^2_0}}
{\Big (}(1+{\frac{r_{\rm s0}}{r}}+{\frac{r_{\text s0}^2}{4r^2}}+
{\frac{r^2}{{\Xi}_0^2}}+{\cdots})dr^2+r^2(1+
O({\frac{r_{\rm s0}^3}{r^3}}))d{\Omega}^2{\Big )}.
\ena
This expression represents approximately the gravitational field outside an 
isolated, approximately metrically stationary, axially symmetric spinning 
source made of perfect fluid (obeying an equation of state of the form 
$p{\propto}{\varrho}_{\text m}$) to the given accuracy in small quantities. Note 
in particular the presence of a tidal term containing the free parameter 
$J_{\rm 2}$, describing the effect of source deformation due to the rotation. 
This tidal term has a counterpart in Newtonian gravitation. (However, to higher
order there is also a term due to the rotation itself, and this term has no 
counterpart in Newtonian gravitation.) The existence of a tidal term means that
the line element family (43) has built into itself the necessary flexibility to 
approximately represent the gravitational field exterior to a variety of 
sources. That is, since the exact equation of state describing the source is 
not specified, the effect of the rotation on the source and thus its 
quadrupole-moment should not be exactly known either, since this effect depends
on material properties of the source. On the other hand, which is well-known, 
no such flexibility is present in the Kerr metric, meaning that the Kerr metric
can only represent the gravitational field outside a source which material 
properties are of no concern; e.g., a spinning black hole.
\section{The geodetic and Lense-Thirring effects}
To calculate the predicted geodetic and Lense-Thirring effects within
the quasi-metric framework we can use the line element family (43) with some
extra simplifications.

We thus examine the behaviour of a small gyroscope in orbit around an 
approximately metrically stationary, axially symmetric, isolated source. We may
assume that the source is so small that any dependence on the global curvature 
of space can be neglected. Furthermore, we assume that the exterior 
gravitational field of the source is so weak that it can be adequately 
represented by equation (43) with the highest order terms cut out, i.e.,
\eqa
{ds_t}^2&=&-{\Big (}1-{\frac{r_{\rm s0}}{r}}+
J_2{\frac{{\bar {\cal R}}^2r_{\text s0}}{2r^3}}(3{\cos}^2{\theta}-1)+
{\cdots}{\Big )}(dx^0)^2 \nonumber \\
&&-2{\frac{t}{t_0}}{\frac{r_{\text s0}a_0}{r}}
{\sin}^2{\theta}d{\phi}dx^0+{\frac{t^2}{t^2_0}}
{\Big (}(1+{\frac{r_{\rm s0}}{r}}+{\cdots})dr^2+r^2d{\Omega}^2{\Big )}.
\ena
In this section we calculate the predicted geodetic effect. In this case
equation (44) can be simplified even further by assuming that the 
gravitational field is spherically symmetric, i.e., that the spin of the
source can be neglected. We then set $J_{\rm 2}=a_0=0$ in equation (44) and
it takes the form
\eqa
{ds_t}^2=-B(r)(dx^0)^2+{\frac{t^2}{t_0^2}}{\Big (}A(r)dr^2+
r^2d{\Omega}^2{\Big )},
\ena
where $A(r)$ and $B(r)$ are given as series expansions from equation (44)
(with spin parameters neglected). 

Our derivation of the geodetic effect in QMR will be a counterpart to a 
similar calculation valid for GR and presented in [7]. To simplify 
calculations we assume that the gyroscope orbits in the equatorial plane and 
that the orbit is a circle with constant radial coordinate $r=R$. Furthermore 
the gyroscope has spin ${\bf S}_t$ and 4-velocity ${\bf u}_t$. Then the norm 
$S_*{\equiv}{\sqrt{{\bf S}_t{\cdot}{\bf S}_t}}$ is constant along its world 
line and moreover normal to ${\bf u}_t$, i.e.
\eqa
{\bf S}_t{\cdot}{\bf u}_t=0, \qquad \Rightarrow \qquad
S_{(t)0}=-S_{(t)i}{\frac{dx^i}{dx^0}}.
\ena
The equation of motion for the spin ${\bf S}_t$ is the equation of 
parallel transport along its world line in quasi-metric space-time, i.e.
\eqa
{\topstar {\nabla}}_{{\bf u}_t}{\bf S}_t=0, \qquad \Rightarrow \qquad
{\frac{dS^{\mu}_{(t)}}{d{\tau}_t}}=-{\topstar {\Gamma}}^{\mu}_{{\lambda}{\nu}}
S^{\lambda}_{(t)}u^{\nu}_{(t)}-{\topstar {\Gamma}}^{\mu}_{{\lambda}t}
S^{\lambda}_{(t)}{\frac{dt}{d{\tau}_t}}.
\ena
Next we define the angular velocity of the gyroscope. This is given by
${\Omega}_t{\equiv}{\frac{d{\phi}}{dt}}=c{\frac{d{\phi}}{dx^0}}$. Thus
$u_{(t)}^{\phi}{\equiv}{\frac{d{\phi}}{d{\tau}_t}}=c^{-1}{\Omega}_t
{\frac{dx^0}{d{\tau}_t}}$. Now a constant of motion $J$ for the orbit is given 
by the equation [3, 4] (using the notation 
$'{\equiv}{\frac{\partial}{{\partial}r}}$)
\eqa
{\frac{t}{t_0}}R^2{\Omega}_t=B(R)Jc, \qquad J={\frac{B'(R)R^3}{2B^2(R)}},
\ena
where the last expression is shown in [8]. Equations (48) then yield
\eqa
{\Omega}_t={\frac{t_0}{t}}{\sqrt{\frac{B'(R)}{2R}}}c.
\ena
Also, from the fact that ${\bf u}_t{\cdot}{\bf u}_t=-c^2$, we find
\eqa
u_{(t)}^0={\frac{dx^0}{d{\tau}_t}}={\frac{c}{\sqrt{B(R)-
{\frac{1}{2}}B'(R)R}}},
\ena
and using equation (50), equation (46) yields
\eqa
S_{(t)}^0={\frac{t}{t_0}}B^{-1}(R){\sqrt{{\frac{1}{2}}B'(R)R^3}}
S_{(t)}^{\phi}.
\ena
We now insert the expressions found above into equation (47).
(The relevant connection coefficients can be found in [3] or [4].)
Equation (47) then yields a set of 2 coupled, first order ordinary
differential equations of the form
\eqa
{\frac{d}{dt}}{\Big [}{\frac{t}{t_0}}S^r_{(t)}{\Big ]}=f(R)S^{\phi}_{(t)}, 
\qquad
{\frac{d}{dt}}{\Big [}{\frac{t}{t_0}}S^{\phi}_{(t)}{\Big ]}=-g(R)S^r_{(t)},
\ena
where the functions $f(R)$, $g(R)$ are given by
\eqa
f(R){\equiv}{\frac{c}{A(R)}}{\Big (}{\sqrt{{\frac{1}{2}}B'(R)R}}-
B^{-1}(R){\sqrt{{\frac{1}{8}}B'^3(R)R^3}}{\Big )}, \qquad
g(R){\equiv}{\sqrt{\frac{B'(R)}{2R^3}}}c.
\ena
A solution of the system (52) can be found by computer. Assuming that
${\bf S}_t$ points in the (positive) radial direction at epoch $t_0$, the
solution of equations (52) reads
\eqa
S_{(t)}^r={\frac{t_0}{t}}S_*A^{-1/2}(R){\cos}{\Big [}{\omega}_{\rm S}t_0
{\ln}({\frac{t}{t_0}}){\Big ]},
\ena
\eqa
S_{(t)}^{\phi}=-{\frac{t_0}{t}}
{\frac{S_*{\Omega}_{t_0}}{{\sqrt{A(R)}}R{\omega}_{\rm S}}}
{\sin}{\Big [}{\omega}_{\rm S}t_0{\ln}({\frac{t}{t_0}}){\Big ]},
\ena 
where
\eqa
{\omega}_{\rm S}{\equiv}{\sqrt{f(R)g(R)}}=A^{-1/2}(R)
{\sqrt{1-{\frac{B'(R)R}{2B(R)}}}}{\Omega}_{t_0}.
\ena
After one complete orbit of the gyroscope $t=t_0+{\frac{2{\pi}}{{\Omega}_t}}$,
and the angle between ${\bf S}_t$ and a unit vector ${\bf e}_r$ in the
radial direction is given by
\eqa
{\alpha}={\arccos}({\frac{{\bf S}_t}{S_*}}{\cdot}{\bf e}_r)=
{\omega}_{\rm S}t_0{\ln}[1+{\frac{2{\pi}}{{\Omega}_tt_0}}].
\ena
The difference ${\Delta}{\phi}$ between a complete circle of $2{\pi}$ radians 
and the angular advancement of ${\bf S}_t$ will then be (with $r_{\rm s0}
{\approx}{\frac{2MG_{\rm N}}{c^2}}$, where $G_{\rm N}$ is Newton's constant)
\eqa
{\Delta}{\phi}=2{\pi}-{\alpha}=2{\pi}{\Big [}{\frac{3r_{\rm s0}}{4R}}-
{\frac{{\pi}R}{{\Xi}_0}}{\sqrt{{\frac{2R}{r_{\rm s0}}}}}+{\cdots}{\Big ]}
{\approx}{\frac{3{\pi}M_{t_0}G_{\rm N}}{c^2R}}-{\frac{2{\pi}^2}{t_0}}
{\sqrt{\frac{R^3}{M_{t_0}G_{\rm N}}}}+{\cdots}.
\ena
We see that there is a quasi-metric correction term in addition to the usual GR
result. Unfortunately, the difference amounts only to about 
$-5{\times}{10^{-5}}''$ per year for a satellite orbiting the Earth, 
i.e. the predicted correction is too small by a factor about ten to be 
detectable by Gravity Probe B.

One may also calculate the Lense-Thirring effect for a gyroscope in
polar orbit, using equation (44). But except from the variable scale
factor, the off-diagonal term in (44) is the same as for the Kerr metric.
Any correction term to the Lense-Thirring effect should therefore depend on 
the inverse age of the Universe and thus be far too small to be detectable.
But note that, for a gyroscope in orbit around the Earth, there is also an
extra contribution term of the type shown in equation (58), to the geodetic 
effect coming from the Earth's orbit around the Sun. Numerically this
correction term is similar to the correction term found above.
\section{Conclusion}
In GR, very many exterior and interior solutions of Einstein's equations are 
possible in principle for axisymmetric stationary systems. The problem is 
to find physically reasonable solutions where the exterior and interior
solutions join smoothly to form an asymptotically flat, global solution. Such 
solutions would be candidates for modelling isolated spinning stars. And 
although no exact solution having the desired properties has been found so 
far, accurate analytical approximations exist, and also numerical solutions of 
the full Einstein equations, see, e.g., [9].

The quasi-metric counterpart to axially symmetric, stationary systems in GR, 
is metrically stationary, axially symmetric systems. This research subject is 
largely unexplored. However, this paper contains some basic results for such 
systems. That is, we have set up the relevant equations for a metrically 
stationary, axially symmetric isolated source within the quasi-metric 
framework, both interior and exterior to the source. Trying to find the first 
few terms of a candidate trial series solution for the exterior part, it was 
found, unexpectedly, that such a solution does not exist. This means that 
{\em exact, metrically stationary, axially symmetric systems do not exist in 
quasi-metric gravity}. Thus all rotating axisymmetric systems must be 
nonstationary according to quasi-metric gravity.

However, for weak gravitational fields and slowly rotating sources, 
axisymmetric systems may be approximated as metrically stationary to very good
accuracy, and an approximate series solution may be found to be valid up to a 
certain order in small quantities. Furthermore, a comparison of this 
approximate series solution to the Kerr metric (to the relevant order in small 
quantities) may be illuminating. The biggest difference between the approximate
series solution and the Kerr metric is the presence of a term containing the 
free parameter $J_{\rm 2}$ representing the quadrupole-moment of the source. 
Such a free parameter is necessary to ensure sufficient flexibility so that the
solution does not unduly constrain the nature of the source. On the other hand,
the multipole moments of the Kerr metric are fixed. This means that, unlike the 
Kerr metric, the approximate metric family found in this paper may 
approximately represent the gravitational field exterior to a variety of 
sources which equation of state satisfies $p{\propto}{\varrho}_{\rm m}$. 
Besides, in the limit where the rotation of the source vanishes, the 
approximately metrically stationary, axially symmetric approximate solution 
becomes identical to the spherically symmetric, metrically static exterior 
solution found in [4] (to the given accuracy). This means that the source's 
quadrupole-moment vanishes in the limit of no rotation and that the source 
should be unable to support shear forces. Thus for such a source, its 
quadrupole-moment is purely due to rotational deformation. It is also possible 
that part of the source's quadrupole-moment is static, in which case the source
cannot be made of perfect fluid. One possible limit of no rotation is then 
given from equation (31).
\\ [6mm]
{\bf References} \\ [2mm]
{\bf [1]} N. Stergioulas, {\em Living Reviews in Relativity} {\bf 6}, 3 
(2003). \\
{\bf [2]} D. {\O}stvang, {\em Grav. \& Cosmol.} {\bf 11}, 205 (2005)
(gr-qc/0112025). \\
{\bf [3]} D. {\O}stvang, {\em Doctoral Thesis}, (2001) (gr-qc/0111110). \\
{\bf [4]} D. {\O}stvang, {\em Grav. \& Cosmol.} {\bf 13}, 1 (2007)
(gr-qc/0201097). \\
{\bf [5]} A. Komar, {\em Phys. Rev.} {\bf 113}, 934 (1959). \\
{\bf [6]} R.M. Wald, {\em General Relativity}, The University of Chicago
Press (1984). \\
{\bf [7]} J.B. Hartle, {\em Gravity: an introduction to Einstein's general 
relativity}, \\
{\hspace*{6.1mm}}Addison-Wesley (2003). \\
{\bf [8]} S. Weinberg, {\em Gravitation and Cosmology}, John Wiley ${\&}$ 
Sons Inc. (1972). \\
{\bf [9]} E. Berti, F. White, A. Manipoulou, M. Bruni,
{\em MNRAS} {\bf 358}, 923 (2005). 
\appendix
\renewcommand{\theequation}{\thesection.\arabic{equation}}
\section{Tensor field equations on the FHSs}
\setcounter{equation}{0}
In this appendix, we set up the nontrivial components of the tensor field 
equation (8) for the axially symmetric, metrically stationary case. To begin 
with, it may readily be shown that for all cases where ${\bar K}_t$ vanishes 
identically, equation (8) takes the same form both interior and exterior to 
any sources present. This form is
\eqa
{\frac{1}{{\bar N}_t}}{\cal L}_{{\bar N}_t{\bf {\bar n}}_t}{\bar K}_{(t)ij}
+{\Big [}{\frac{2}{3}}{\bar K}_{(t)ks}{\bar K}_{(t)}^{ks}
-{\frac{1}{(ct{\bar N}_t)^2}}{\Big ]}{\bar h}_{(t)ij}={\tilde H}_{(t)ij} .
\ena
Given equation (20), there are four different nontrivial components of 
equation (A.1), namely
\eqa
{\tilde H}_{(t){\rho}{\rho}}={\frac{1}{2}}{\Big {\{}}-
{\frac{{\bar Y},_{{\rho}{\rho}}}{\bar Y}}-
{\frac{{\bar Z},_{{\rho}{\rho}}}{\bar Z}}+{\frac{2}{\rho}}{\Big (}
{\frac{{\bar X},_{\rho}}{\bar X}}-{\frac{{\bar Y},_{\rho}}{\bar Y}}-
{\frac{{\bar Z},_{\rho}}{\bar Z}}{\Big )}+{\frac{1}{2}}{\Big (}
{\frac{{\bar X},_{\rho}}{\bar X}}+{\frac{{\bar Y},_{\rho}}{\bar Y}}{\Big )}
{\frac{{\bar Y},_{\rho}}{\bar Y}} \nonumber \\
+{\frac{1}{2}}{\Big (}
{\frac{{\bar X},_{\rho}}{\bar X}}+{\frac{{\bar Z},_{\rho}}{\bar Z}}{\Big )}
{\frac{{\bar Z},_{\rho}}{\bar Z}}+
{\frac{1}{{\Xi}_0^2(1-{\frac{{\rho}^2}{{\Xi}_0^2}})}}{\Big [}4-6{\bar X}+
{\rho}{\frac{{\bar Y},_{\rho}}{\bar Y}}+{\rho}{\frac{{\bar Z},_{\rho}}{\bar Z}}
{\Big ]} \nonumber \\
+{\frac{{\bar X}}{{\bar Y}{\rho}^2(1-{\frac{{\rho}^2}{{\Xi}_0^2}})}}
{\Big [}-{\frac{{\bar X},_{{\theta}{\theta}}}{\bar X}}+{\frac{1}{2}}{\Big (}
{\frac{{\bar X},_{\theta}}{\bar X}}+{\frac{{\bar Y},_{\theta}}{\bar Y}}-
{\frac{{\bar Z},_{\theta}}{\bar Z}}{\Big )}{\frac{{\bar X},_{\theta}}{\bar X}}
-{\cot}{\theta}{\frac{{\bar X},_{\theta}}{\bar X}}{\Big ]}
{\Big {\}}} \nonumber \\
=-{\frac{{\bar X}}{{\Xi}_0^2(1-{\frac{{\rho}^2}{{\Xi}_0^2}})}}
-{\frac{1}{3}}{\bar Z}{\rho}^2{\sin}^2{\theta}{\Big [}2({\bar V},_{\rho})^2
-{\frac{{\bar X}}{{\bar Y}{\rho}^2(1-{\frac{{\rho}^2}{{\Xi}_0^2}})}}
({\bar V},_{\theta})^2{\Big ]},
\ena
\eqa
{\tilde H}_{(t){\theta}{\theta}}={\frac{1}{2}}
{\Big {\{}}{\frac{{\rho}^2{\bar Y}}{{\bar X}}}
{\Big (}1-{\frac{{\rho}^2}{{\Xi}_0^2}}{\Big )}{\Big [}
-{\frac{{\bar Y},_{{\rho}{\rho}}}{\bar Y}}+{\frac{1}{\rho}}{\Big (}
{\frac{{\bar X},_{\rho}}{\bar X}}-3{\frac{{\bar Y},_{\rho}}{\bar Y}}-
{\frac{{\bar Z},_{\rho}}{\bar Z}}{\Big )} \nonumber \\
+{\frac{1}{2}}{\Big (}
{\frac{{\bar X},_{\rho}}{\bar X}}+{\frac{{\bar Y},_{\rho}}{\bar Y}}-
{\frac{{\bar Z},_{\rho}}{\bar Z}}{\Big )}
{\frac{{\bar Y},_{\rho}}{\bar Y}}-{\frac{2}{{\rho}^2}}{\Big ]}
+{\frac{\bar Y}{\bar X}}{\frac{{\rho}^2}{{\Xi}_0^2}}{\Big (}2-6{\bar X}+
{\rho}{\frac{{\bar Y},_{\rho}}{\bar Y}}{\Big )}-
{\frac{{\bar X},_{{\theta}{\theta}}}{\bar X}}-
{\frac{{\bar Z},_{{\theta}{\theta}}}{\bar Z}} \nonumber \\
+{\frac{1}{2}}
{\Big (}{\frac{{\bar X},_{{\theta}}}{\bar X}}{\Big )}^2+{\frac{1}{2}}
{\Big (}{\frac{{\bar Z},_{{\theta}}}{\bar Z}}{\Big )}^2+{\frac{1}{2}}
{\Big (}{\frac{{\bar X},_{\theta}}{\bar X}}+
{\frac{{\bar Z},_{\theta}}{\bar Z}}{\Big )}{\frac{{\bar Y},_{\theta}}{\bar Y}}
+{\cot}{\theta}{\Big (}{\frac{{\bar Y},_{\theta}}{\bar Y}}-
2{\frac{{\bar Z},_{\theta}}{\bar Z}}{\Big )}+2{\Big {\}}} \nonumber \\
=-{\bar Y}{\frac{{\rho}^2}{{\Xi}_0^2}}
+{\frac{1}{3}}{\bar Z}{\rho}^2{\sin}^2{\theta}{\Big [}
{\frac{\bar Y}{\bar X}}{\rho}^2(1-{\frac{{\rho}^2}{{\Xi}_0^2}})
({\bar V},_{\rho})^2-2({\bar V},_{\theta})^2{\Big ]},
\ena
\eqa
{\tilde H}_{(t){\phi}{\phi}}={\frac{1}{2}}{\bar Z}{\sin}^2{\theta}
{\Big {\{}}{\frac{{\rho}^2}{{\bar X}}}
{\Big (}1-{\frac{{\rho}^2}{{\Xi}_0^2}}{\Big )}{\Big [}
-{\frac{{\bar Z},_{{\rho}{\rho}}}{\bar Z}}+{\frac{1}{\rho}}{\Big (}
{\frac{{\bar X},_{\rho}}{\bar X}}-{\frac{{\bar Y},_{\rho}}{\bar Y}}-
3{\frac{{\bar Z},_{\rho}}{\bar Z}}{\Big )} \nonumber \\
+{\frac{1}{2}}{\Big (}{\frac{{\bar X},_{\rho}}{\bar X}}-
{\frac{{\bar Y},_{\rho}}{\bar Y}}{\Big )}{\frac{{\bar Z},_{\rho}}{\bar Z}}
+{\frac{1}{2}}{\Big (}{\frac{{\bar Z},_{{\rho}}}{\bar Z}}{\Big )}^2{\Big ]}
+{\frac{2}{{\bar X}{\bar Y}}}{\Big (}{\bar X}-{\bar Y}{\Big )}+
{\frac{1}{\bar X}}{\frac{{\rho}^2}{{\Xi}_0^2}}{\Big (}4-6{\bar X}{\Big )}
\nonumber \\
+{\frac{1}{\bar Y}}{\Big [}-{\frac{{\bar Z},_{{\theta}{\theta}}}{\bar Z}}
+{\frac{1}{2}}{\Big (}{\frac{{\bar Z},_{{\theta}}}{\bar Z}}{\Big )}^2
+{\frac{1}{2}}{\Big (}{\frac{{\bar Y},_{\theta}}{\bar Y}}-
{\frac{{\bar X},_{\theta}}{\bar X}}{\Big )}{\frac{{\bar Z},_{\theta}}{\bar Z}}
+{\cot}{\theta}{\Big (}{\frac{{\bar Y},_{\theta}}{\bar Y}}-
{\frac{{\bar X},_{\theta}}{\bar X}}-
2{\frac{{\bar Z},_{\theta}}{\bar Z}}{\Big )}{\Big ]}{\Big {\}}} \nonumber \\
=-{\bar Z}{\sin}^2{\theta}{\frac{{\rho}^2}{{\Xi}_0^2}}
+{\frac{1}{3}}{\bar Z}{\rho}^2{\sin}^4{\theta}{\Big [}
{\frac{\bar Z}{\bar X}}{\rho}^2(1-{\frac{{\rho}^2}{{\Xi}_0^2}})
({\bar V},_{\rho})^2+{\frac{\bar Z}{\bar Y}}({\bar V},_{\theta})^2{\Big ]},
\ena
\eqa
{\tilde H}_{(t){\rho}{\theta}}&=&{\frac{1}{2}}{\Big {\{}}
-{\frac{{\bar Z},_{{\rho}{\theta}}}{\bar Z}}+{\frac{1}{2}}
{\Big (}{\frac{{\bar X},_{\theta}}{\bar X}}+
{\frac{{\bar Z},_{\theta}}{\bar Z}}{\Big )}{\frac{{\bar Z},_{\rho}}{\bar Z}}
+{\frac{1}{2}}{\frac{{\bar Y},_{\rho}{\bar Z},_{\theta}}{{\bar Y}{\bar Z}}}
\nonumber \\
&&+{\frac{1}{\rho}}{\frac{{\bar X},_{\theta}}{\bar X}}+
{\cot}{\theta}{\Big (}{\frac{{\bar Y},_{\rho}}{\bar Y}}-
{\frac{{\bar Z},_{\rho}}{\bar Z}}{\Big )}{\Big {\}}}=
-{\bar Z}{\rho}^2{\sin}^2{\theta}({\bar V},_{\rho})({\bar V},_{\theta}).
\ena
For a given function ${\bar V}({\rho},{\theta})$, equations (A.2)-(A.5) 
represent four coupled equations for the three unknown quantities ${\bar X}$,
${\bar Y}$ and ${\bar Z}$. It would thus seem that this system is 
overdetermined.

To find an approximate perturbative solution as the first few terms in a series
expansion in small quantities, first set ${\bar X}={\bar Y}={\bar Z}=1$ in
equations (29) and (30) and solve those equations for ${\bar B}$ and ${\bar V}$
to lowest order in small quantities. Then insert the obtained lowest order
expression for ${\bar V}$ into equations (A.2)-(A.5) and seek for the lowest 
order solution. Unfortunately, it is found from this procedure that the 
equation system (A.2)-(A.5) has no solution given the found expression for 
${\bar V}({\rho},{\theta})$. One might think that this result follows from the
fact that the system is overdetermined, so it might seem like a good idea to
introduce an extra term
$2{\frac{t^2}{t_0^2}}{\bar W}({\rho},{\theta})d{\rho}d{\theta}$ in equation 
(20). However, this possibility will considerably increase the complexity of 
equations (A.2)-(A.5) and in the end it does not seem to help (any trial series
solution should have cylindrical symmetry).
\end{document}